# ShopList: Programming PDA applications for Windows Mobile using C#

**Student Daniela Ilea, Assoc. Prof. Ph.D.Eng. Dan L. Lacrama**
"Tibiscus" University of Timisoara, Romania

**ABSTRACT.** This paper is focused on a C# and Sql Server Mobile 2005 application to keep evidence of a shop list. The purpose of the application is to offer to the user an easier way to manage his shopping options.

## 1.  Generalities

During the last years wireless technology has knew an extraordinary evolution: devices like PDA or SmartPhone became more and more impressive regarding the technogy involved. PDA terminology comes from Personal Digital Assistant, used for the first time in 1992, but the technology is describe since the 70s, as "very performant computers", "organizers", or, finally: palmtops.

Many operating systems have been developed : WindowsMobile, Windows CE, Palm OS, BlackBerry OS, Symbian, and a few other operating systems, based on Linux (Freeware) : Maemo, GPE, OPIE. Every operating system has its own strong points, but the most used is Windows Mobile. The Gartner Institute realized an market reserche about the most used mobile operating systems, and the results look like this:

- Windows Mobile        (49.2 %)
- RIM Black Berry       (25%)
- Palm OS               (14.9%)
- Symbian
- Other

The application is using Visual Studio 2005 , C# language and Sql Server Mobile Edition 2005. Why this combination?

- Because full integration of SQL Server 2005 Mobile Edition with SQL Server 2005 and Visual Studio 2005 provides a platform for developers to rapidly build applications that extend enterprise data management capabilities to mobile devices.
- Sql Server 2005 Mobile Edition is freeware for use and deploy.
- The environment is very friendly, offering an emulator and an easy way to deliver the application directly on the mobile device

Microsoft SQL Server 2005 Mobile Edition 3.0 (SQL Server Mobile) supports two methods of exchanging data with a SQL Server database:

- Merge replication, which provides a robust full-featured solution that allows a mobile application to make autonomous changes to replicated data, and at a later time, merge those changes with a Microsoft SQL Server database, and resolve conflicts when necessary.
- Remote data access (RDA) provides a simple way for a mobile application to access (pull) and send (push) data to and from a remote Microsoft SQL Server database table and a local SQL Server Mobile database table. RDA can also be used to issue SQL commands on a server running SQL Server.

Microsoft Visual Studio is Microsoft's flagship software development product for computer programmers. It centers on an integrated development environment which lets programmers create standalone applications, web sites, web applications, and web services that run on any platforms supported by Microsoft's .NET Framework (for all versions after 6). Supported platforms include Microsoft Windows servers and workstations, PocketPC, Smart phones, and World Wide Web browsers.

The requirements for this application are:

- The device needs Windows Mobile as operating system
- . Net framework 2.0, freeware
- Sql Server Mobile Edition 2005, freeware

The Microsoft .NET Framework is a software component that can be added to or is included with the Microsoft Windows operating system. It provides a large body of pre-coded solutions to common program requirements, and manages the execution of programs written specifically for the framework. The .NET Framework is a key Microsoft offering, and is intended to be used by most new applications created for the Windows platform.

## 2. Shop List Application

The application has four sections:
- Category: allows the user to enter product category, such as: food, clothes, gifts
- Product: allows user to enter products related to a specific category. The user can specify a product as Favorite, so when creating a new shop list, he doesn't need to choose the product from a large list, just Press "Show favorites" and add from a selective list the chosen product.
- New option is creating a new Shopping List. The user can choose a product or add one from the favorites list.
- List: shows the current shop list, with all products colored in red. When the user check a product from the list, meaning "Ok, I bought this product", the product is colored in green

The database is included in application, local, for performance considerations. It includes three tables: Categories, Products and List, related to each other.

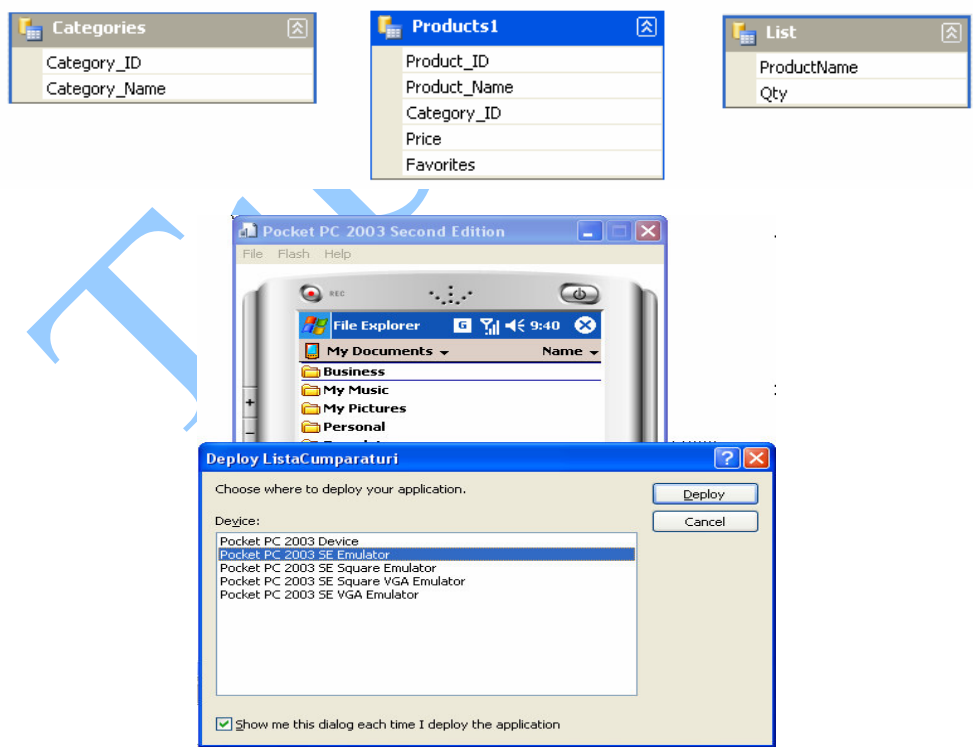

Building the application, we have two options:
1. Build on emulator, good to use during the development process until the application is ready to be delivered
2. Build on the device directly. This is the most simple way to deliver the application. Visual Studio detects all dependencies (such as .net framework, or the libraries used ) and deliver it on the machine.

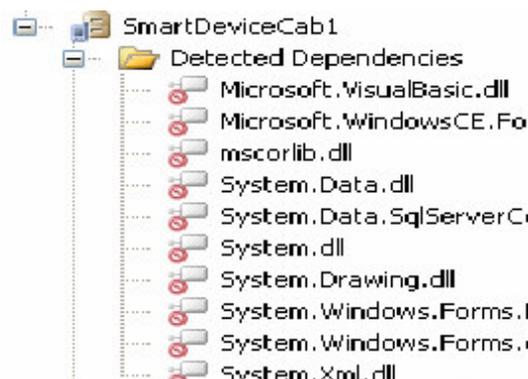

**2.1 Classes in Shop List Applications**
To develop this application I needed to use a few of system classes. Next to common classes, such as System.Windows.Forms, or System.IO, I also used other classes:
- System.Environement: The Environment class defined in the System namespace allows developers to get the information about the current environment and platform: current directory, machine name, User Name, OS version.
- System.Data.SqlServerCe : to communicate with the database. This class contains :
  o SqlCeConnection.
  o SqlCeDataAdapter : SqlCeDataAdapter Class represents a set of data commands and a database connection that are used to fill the Dataset and update the data source.
  o SqlCeCommand: Represents an SQL statement to execute against a data source.

My own classes used in the application are:
- DB.cs: is used to realize he connection to the database. Because the database is not located on a server, but included and delivered with the application, I used System.Reflection to locate the application and, in the same directory as the executable, the database.

StringBuilder s = new StringBuilder(@"Data Source = " + Path.GetDirectoryName(System.Reflection.Assembly.GetExecutingAssembly().GetName().CodeBase) + @"\ShopList.sdf; Password ="");
cn.ConnectionString = s.ToString();

- MainForm.cs is the main form of the application. It contains the tab with all the application's options

*Adding category*

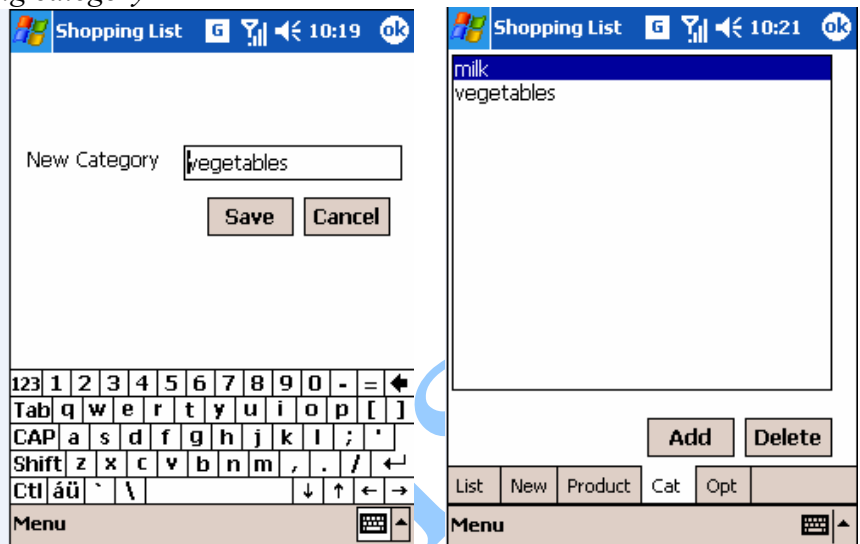

*Create a new shop list from favorites products*

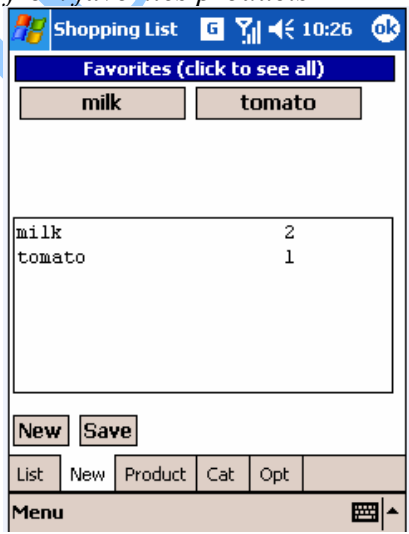

- FrmProduct contains all Product related data: select from database, insert or update in database

```
System.Data.SqlServerCe.SqlCeCommand cmd = new
System.Data.SqlServerCe.SqlCeCommand();
cmd.Connection = DB.GetCn;
cmd.CommandType = CommandType.Text;
cmd.CommandText = "Insert into Products(Product_Name, Price)
        values ('" + TxtProdName.Text + "', " + TxtPrice.Text + ")";
cmd.ExecuteNonQuery();
```

*Adding product*

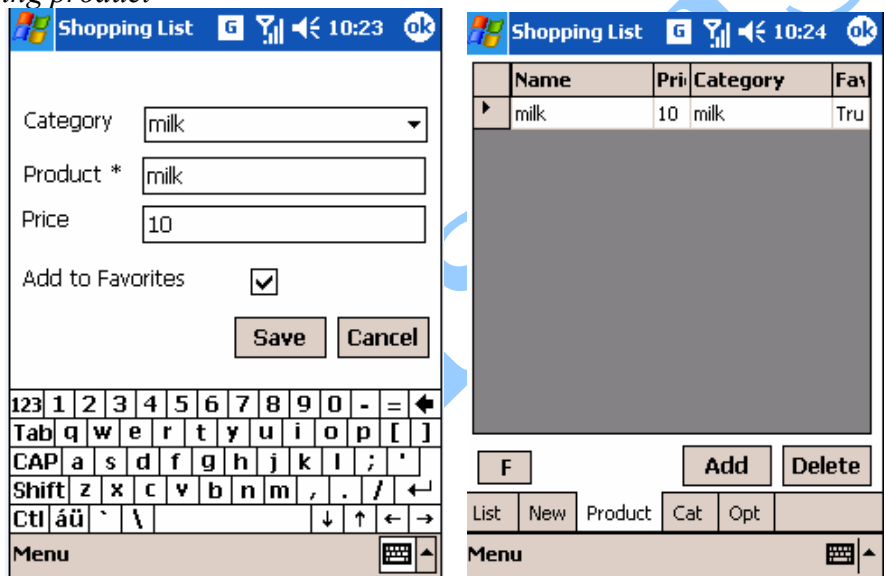

- DebugUtils.cs: contains some test executed to measure the performance, unrelated directly to the design of the application. Here is an example of measuring the time needed to an object to appear:

```
public static void DisplayMeasureResult( string s )
    {
       endTickCount = Environment.TickCount;
       timeTaken = endTickCount - startTickCount;
       System.Windows.Forms.MessageBox.Show(s+ " took: " +
timeTaken.ToString() + "ms");
    }
```

*Shop List*

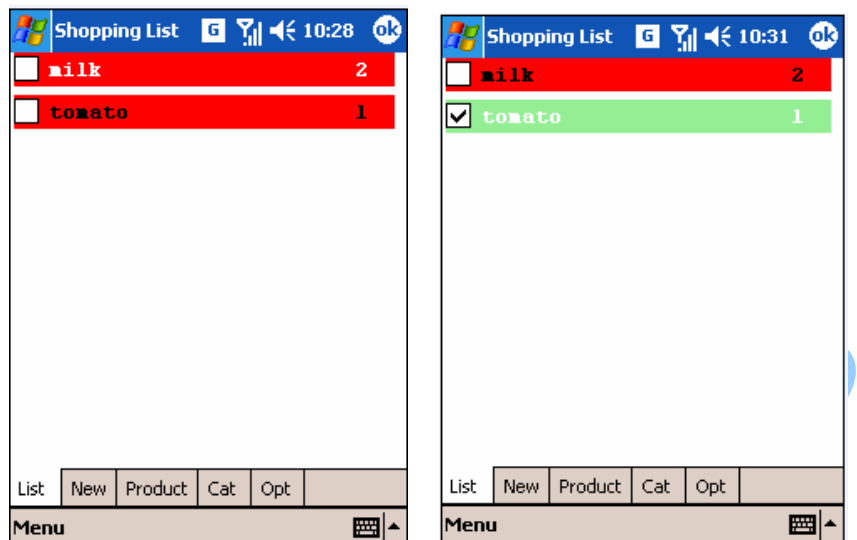

## Conclusions

Visual Studio 2005 is a very suitable environment to develop mobile device applications, offering the environment and the database. The environment is very friendly, containing an emulator and an easy and fast way to deploy the application.